\documentclass[prb, twocolumn, superscriptaddress]{revtex4}
\usepackage{amsmath}
\usepackage{amsfonts}
\usepackage{amssymb}
\usepackage{graphicx}

\begin{document}
\title{A unified analytical model for extraordinary transmission in subwavelength metallic gratings}
\author{Ilai Schwarz}
\affiliation{Racah Institute of Physics, The Hebrew University, Jerusalem 91904 Israel}
\author{Nitzan Livneh}
\affiliation{Applied Physics Department, The Benin School of computer sciences and engineering, The Hebrew University, Jerusalem 91904 Israel}
\author{Ronen Rapaport}
\affiliation{Racah Institute of Physics, The Hebrew University, Jerusalem 91904 Israel}
\affiliation{Applied Physics Department, The Benin School of computer sciences and engineering, The Hebrew University, Jerusalem 91904 Israel}

\begin{abstract}
We present an intuitive analytical model for extraordinary optical transmission (EOT) through corrugated metallic films. In the framework of this model, EOT emerges from standing wave resonances of the different diffraction orders, and is an inherent property of periodic structures, which does not require specific polarization dependent properties such as surface plasmons. The model correctly predicts the conditions for the EOT resonances in various geometrical configurations, in both TE and TM polarizations,and in the subwavelength and non-subwavelength spectral regimes, using the same underlying mechanism.
\end{abstract}
\maketitle

\section{Introduction \label{intro}}

The electromagnetic response of subwavelength metallic corrugated structures give rise to many physically interesting, potentially useful, and sometime initially counterintuitive phenomena. Perhaps the most widely known is that of resonant extraordinary optical transmission (EOT) in arrayed subwavelength holes or slits in metallic sheets \cite{ebbesen_extraordinary_1998}. In subwavelength slit arrays, EOT is defined as light transmission superseding the geometrical ratio of the open, transparent slits area and the opaque, metal surface area. Carefully engineered subwavelength slit arrays were shown to exhibit other interesting qualities, such as beaming and focusing of light \cite{nanfang_yu_plasmonics_2010}), and large local field amplifications \cite{garcia-vidal_transmission_2002} among others.
Such structures therefore has been suggested as functional parts for various device realizations \cite{zhang_band-selective_2008, chien_coupled_2007}. 

 While numerical calculations and semi-analytical models accurately predicted the emergence of EOT in 1D arrays of subwavelength metallic gratings \cite{porto_transmission_1999, treacy_dynamical_2002, lalanne_one-mode_2000, shen_properties_2004}, the underlying physical mechanism has been under scientific debate for several years\cite{porto_transmission_1999,cao_negative_2002,garcia-vidal_transmission_2002,pendry_mimicking_2004, treacy_dynamical_2002}. Most explanations involved EOT for incoming light in TM polarization (the magnetic field component of the incoming light is parallel to the slits), and invoked, in one way or the other, resonant excitation of modified forms of surface waves. For 1D slits, TM polarized light always has an electric field component perpendicular to the slits, and can therefore resonantly excite surface waves - surface plasmons (SPP)\cite{garcia-vidal_transmission_2002}. These phenomena were thought to be essential for EOT, though different mechanisms such as the full dynamical diffraction theory\cite{treacy_dynamical_2002}, and cavity-like EOT were also suggested as possible explanations for emergence of EOT\cite{porto_transmission_1999,cao_negative_2002}.

In 2006 E. Moreno et al. \cite{moreno_extraordinary_2006} showed theoretically that EOT is possible for TE polarization as well, where surface excitations are not allowed, and therefore such EOT is essentially a plasmonless EOT. They showed that such plasmonless EOT can be realized only by laying a thin dielectric layer on top of the metallic grating. In their explanation, the reason for the existence of such a phenomenon is based on a coupling of the incoming light to slab waveguide modes in the thin dielectric layer, which take the role of surface waves as the resonant mediator for the light transmission. Such unique structures and related effects where actually both calculated and experimentally realized previously (in perhaps a different context) by Rosenblatt et al. \cite{rosenblatt_resonant_1997}, which termed them grating waveguide structures. Several later works have investigated various modifications of such grating waveguide structure designs. However, recent works have shown the emergence of EOT in the TE polarization without the extra dielectric layer \cite{crouse_polarization_2007,lochbihler_enhanced_2009}, so apparently there is no mechanism involved which could take the role of the SPPs.

Notably, a comprehensive study of both TE and TM EOT on the same footing and an intuitive model that explains both phenomena in a unified, simple way is yet unavailable as far as we know. Such a description, if exists, can lead to a clearer, more intuitive understanding of the underlying mechanism of EOT in such structures, and can also be utilized for designing such structures for various applications.

Therefore, the purpose of this article is not to develop a different numerical method by which a rigorous simulation of the transmission spectra can be calculated, but rather to propose the simplest analytical model which will still be able to predict the emergence of EOT. Such a model has to be in good agreement with the experiments and rigorous numerical models. Furthermore, the model has to be appropriate for a wide variety of configurations without modifying its basic building blocks, and should give a clear picture of the underlying physical mechanism behind the apriori different EOT situations.

Here we present an approximated model in which the EOT resonances occur when incoming light diffracts to various bragg modes due to the periodic metallic structure. For these bragg modes, and specifically for the first bragg mode, the metal grating can be treated as an \emph{effective dielectric medium}. For some wavelengths, the first bragg mode can form a standing wave between two effective edges of the structure. Whenever such resonant standing waves occur, the structure behaves as a Fabri-Perot like ethalon and therefore a very high transmission. We show that this approximated picture can be used to quantitatively describe, on the same footing, EOT in both the TE and TM polarizations of the incoming light, with or without finite thickness dielectric layers surrounding the metallic grating, thus giving a unified intuitive picture of the mechanism behind EOT in such structures.

\begin{figure}
\includegraphics[width=60mm]{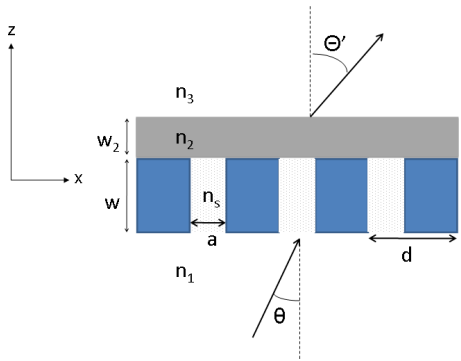}
\caption{\label{fig:configuration} A general metallic grating cross section configuration. The incident plane wave vector, as well as the transmitted wave vector are represented by the arrows. All the relevant physical parameters are explained in the text.}
\end{figure}

\section{Effective Bragg-Cavity Model: Derivation \label{EBC}}

The structure geometry under consideration and all the relevant geometrical, optical, and material parameters are shown in Fig.~\ref{fig:configuration}. Here the incoming plane wave has a positive $k_z$, $d$ is the periodicity of the grating, $a$ the slit width, $w$ is the grating thickness. $n_1, n_3$ are the refractive index of the infinite dielectric layers before and after the grating, and $n_s$ is the refractive index inside the slits. In some of the configurations, there will be an extra thin dielectric layer with the refractive index $n_2$, with a  thickness $w_2$, of the same order of magnitude as the grating thickness $w$. 

\begin{figure}
\includegraphics[width=80mm]{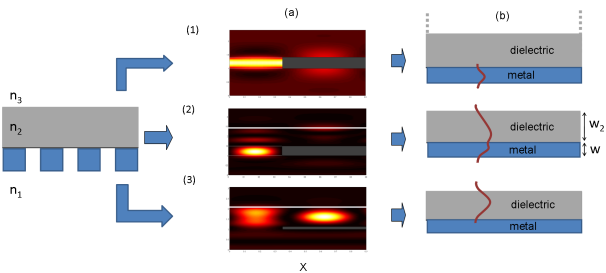}
\caption{\label{fig:model} A schematic illustration of the EBC model. (a) Near field intensity in a unit cell at a wavelength corresponding to an EOT maximum in three different configurations, as explained in the text and in table \ref{table1}. (b) The corresponding model schematics. The same model applies to all the configurations, the only difference is the area in which the standing wave appears in the structure.}
\end{figure}

We will now show how an approximation for the EOT maxima can be found by the condition of a standing wave in the subwavelength corrugated structure for the first bragg order. This will lead to a mapping of this structure to a similar configuration with an effective homogeneous dielectric layer replacing the grating as shown in fig.~\ref{fig:model}, and to the Effective Bragg-Cavity model. We start our derivation  from the Maxwell equations in an inhomogeneous medium,

\begin{equation}\label{maxE}
    \nabla\times \left[\frac{1}{\mu(\textbf{r})}\nabla\times \textbf{E(r)} \right] - k^2\epsilon\textbf{(r)E(r)} = 0.
 \end{equation}

\begin{equation}\label{maxH}
    \nabla\times \left[\frac{1}{\epsilon\textbf{(r)}}\nabla\times \textbf{H(r)} \right] - k^2\mu\textbf{(r)H(r)} = 0.
 \end{equation}
Where $ \epsilon $ is the dielectric constant and $ \mu $ is the magnetic permeability. The vacuum wave vector of the plane wave with a wavelength $ \lambda $, incident on the grating is $ k_0 = 2\pi / \lambda $.

Using these equations, we can find the eigenvalues of Eq.~\ref{maxH} in each of the four layers depicted in Fig~\ref{fig:configuration} separately.
In the homogeneous dielectric materials before and after the grating, Eqs.~\ref{maxE},~\ref{maxH} reduce to
$$ \triangle \textbf{E(r)} + k^2\epsilon\mu \textbf{E}(\textbf{r}) = 0. $$
$$ \triangle \textbf{H(r)} + k^2\epsilon\mu \textbf{H(r)} = 0. $$
Producing the known wave equations in dielectric media.

Let the incident plane wave be in the TM polarization. assuming a one-dimensional array of slits, periodic along the x axis, with $\mu(r) = 1$ everywhere, the eigenfunctions of Eq.~\ref{maxH}, and therefore, for the magnetic field inside the metal grating will be of the form of Bloch waves \cite{treacy_dynamical_2002}:
$$ H^{(j)}(r) = \sum_m{H_{mj}e^{i[(k_x+gm)x + k_z^{(j)}z]}\hat{y}}, $$
where $ g = \frac{2\pi}{d} $, $ k_x $ is the same as that of the incident electromagnetic wave, and $ k_z^{(j)} $ will be found from Eq.~\ref{maxH}.
The total magnetic field is given by the sum of each Bloch-wave excitation
\begin{equation}\label{blochH}
H(r) = \sum_j{\psi_{(j)}}\sum_m{H_{mj}e^{i[(k_x+gm)x + k_z^{(j)}z]}\hat{y}},
\end{equation}
here $\psi_{(j)}$ is the exitation of the j-th eigen mode inside the grating. In the dielectric layers before and after the grating, indexed by 1 and 3 respectively,
$$ H_{1,3}(r) = \sum_m{A_m^{1,3}e^{i[(k_x+gm)x + k_z^{1,3}z]}\hat{y}}, $$
with $k_z$ given by $k_z^{1,3} = \sqrt{{(k_{1,3})}^2 - {(gm)}^2}$, and $k_{1,3} = k_0n_{1,3}$.

Let us focus on normal incident light (i.e. $\theta=0$ in fig.~\ref{fig:configuration}).
In the subwavelength regime ($\frac{\lambda}{n_s} > 2a$), there is only one propagating mode in the grating, with $k_z^{(j)} = k_z^{prop}=k_s$, $k_s=n_sk$. Using the approximation that only this mode is excited by the incoming plane wave (which means neglecting the evanescent modes inside the metallic grating), Eq.~\ref{blochH} becomes:
\begin{equation}\label{SingleMode}
H(r) = \sum_m{H_me^{i[(k_x+gm)x + k_z^{prop}z]}\hat{y}},
\end{equation}
and solving the full boundary condition problem under this approximation leads to semi-analytical models for transmission \cite{porto_transmission_1999, lalanne_one-mode_2000, shen_mechanism_2005}.

However, even without an exact solution to the equations, one can intuitively identify the cause for the EOT: Under the condition that the wavelength satisfies $\frac{\lambda}{n_{1,3}} > d$, which means $k_{1,3} < g$, there will be only one propagating mode outside of the grating, having $k_x = 0$, while all the modes having $k_x = gm$, where $m\neq0$ are evanescent. However, inside the grating, there will still be propagating modes with $k_x = gm$, because of the periodic bloch function, as is clearly seen in Eq.~\ref{SingleMode}. Hence, these modes, which are evanescent outside the grating, will be confined to the grating \cite{shen_mechanism_2005}. Since the only excited mode in the grating is the propagating mode, having $k_z = k_z^{prop}$ and $k_x=gm$, we can map this problem into a similar one: replacing the metallic grating with a dielectric material whose refractive index is defined as  $n_{eff}^m = \sqrt{{(k_z^{prop})}^2 + (gm)^2}/k$. For normal incidence in the TM polarization, $k_z^{prop} = n_sk_0$. Thus $n_{eff}^m = \sqrt{n_s^2+(m\lambda/d)^2}$. Now the boundary matching condition for a given bragg mode in the metallic grating is the same as the boundary condition for a waveguide mode with $n = n_{eff}^m$. Thus, the standard slab waveguide transverse resonance condition
\begin{equation}\label{EOTConditionTM}
2k_z^{prop}w - 2\phi_{12} - 2\phi_{23} = 2\pi l,
\end{equation}
will give us the values of $k$ \emph{which produces the standing wave inside the grating layer}, which will be denoted by $k_0$.
For $m=1$, $\phi_{12} = \tan^{-1}(\hat{\gamma}/k_z^{prop})$, $\phi_{23} = \tan^{-1}(\hat{\delta}/k_z^{prop})$, $\hat{\gamma} = (n_{eff}/n_1)^2\sqrt{g^2 - (n_1k_0)^2}, \hat{\delta} = (n_{eff}/n_3)^2\sqrt{g^2 - (n_3k_0)^2}$ and $n_1, n_3$ are the refractive index before and after the grating respectively. $l$ is a non negative integer. For the case where $n_s=n_1=n_3$, we get
\begin{equation}
\label{phi123}
\phi_{12} = \phi_{23} = tan^{-1}\left((\chi^2+1)\sqrt{\chi^2-1}\right).
\end{equation}
with $\chi = g/(n_sk)$

For the TE polarized plane waves, the derivation is similar, starting from Eq.~\ref{maxE}, and using a similar procedure, we arrive at the slab waveguide transverse resonance condition in the TE polarization:
\begin{equation}\label{EOTConditionTE}
2k_z^{prop}w - 2\phi_{12} - 2\phi_{23} = 2\pi l,
\end{equation}
 With $\phi_{12} = \tan^{-1}(\gamma/k_z^{prop})$, $\phi_{23} = \tan^{-1}(\delta/k_z^{prop})$, $\gamma = \sqrt{g^2 - (n_1k_0)^2}, \delta = \sqrt{g^2 - (n_3k_0)^2}$.

The crucial point is that, as in a Fabri-Perot ethalon, these standing wave conditions will have a visible effect on the transmission. We argue that for $m=0$ these standing waves will cause the forward transmission to be at maximum value, because of constructive interference similar to a Fabri-Perot ethalon. The argument is similar in spirit to the one proposed by Rosenblatt et al.\cite{rosenblatt_resonant_1997} for reflection resonances in grating waveguide structures. Since the waveguide condition is transcedental, an analytical equation showing that the standing wave condition and the EOT condition are the same, is difficult to achieve. However, we will show that for a range of different configurations, with incoming light in the TE or TM polarizations, the wavelength $\lambda_0$, for which an EOT maxima occurs, given by rigorous numerical calculations and the one found by the analytical equation for the standing wave condition are the same. \emph{Therefore, our simplified model is that for there to be EOT, there has to be a standing wave in the $\hat{z}$ direction inside the system for the bragg modes having $m\neq0$}.

In the case where $\frac{a}{d} \sim 0.5$, a good approximation is taking only the first bragg diffraction, i.e. $m=1$, since for ideal metals, $H_m$ will be proportional to the fourier transform of a rectangular box of width $\frac{a}{d}$, which decreases as $sinc(\frac{a}{d})$. (It can be easily shown that this is also the waveguide-like condition found by Shen et al. \cite{shen_mechanism_2005} in the limit $\sum_m sinc(\frac{am}{d})\approx sinc(\frac{a}{d})$). Even for real metals, this approximation still holds, as will be evident from our comparison to rigorous numerical calculations in the next section and from calculating the values of $H_m$ for different real metals (approximated by the drude model). Therefore, it is possible to treat the grating as a dielectric layer with 
 \begin{equation}
 \label{neff}
n_{eff} = \frac{\sqrt{{(k_z^{prop})}^2 + g^2}}{k}
 \end{equation}
(since m=1), and find the wavelength that satisfies the slab waveguide resonance condition of Eqs.~\ref{EOTConditionTM},~\ref{EOTConditionTE}. \emph{This, in essence, is the Effective Bragg-Cavity (EBC) Model.} This simple model predicts correctly the emergence of EOT in a vast variety of 1D configurations, in both TE and TM polarizations, and for the subwavelength and non-subwavelength spectral regimes, \emph{using the same condition}.

Fig.~\ref{fig:model} shows a schematic representation of the model. There are three different cases in which a standing wave can be achieved. Fig.~\ref{fig:model}(a)(1-3) shows the numerically calculated near field intensity for an incoming $\lambda$ at an EOT maximum in different configurations: (1) corresponds to an incident plane wave in either the TM polarization, in which case both $(\lambda/n_s)>2a$ and $(\lambda/n_s)<2a$ are valid, or in the TE polarization, for $(\lambda/n_s)<2a$. Both the dielectric materials $n_1, n_3$ are approximated as having infinite thickness. This is the usual scenario of a bare grating discussed in the literature (specifically, the near field calculation shown coincides with TM polarized light). Fig.~\ref{fig:model}(a)(2) corresponds to the same regimes as Fig.~\ref{fig:model}(a)(1), but with an added dielectric layer $n_2$ with a finite thickness, of the same order as the metallic grating thickness. (3) corresponds to an incoming plane wave in the TE polarization, with $(\lambda/n_s)>2a$, and the dielectric layer $n_2$ having a finite thickness. Fig.~\ref{fig:model}(b)(1-3) shows the model schematics, with the standing wave which corresponds to each of the configurations. All these configurations are summarized in table \ref{table1} . It is clear that the model is the same for all three configurations, and the only difference is the area which confines the standing wave at the EOT resonances.

To show the generality of this simple picture, we note again that our model predicts EOT in the TE polarization as well. Let us then first explain the emergence of EOT in the TE polarization. As stated earlier, for incoming plane waves in the TE polarization, one starts from Eq.~\ref{maxE}, and derives Eq.~\ref{EOTConditionTE} as the EOT resonance condition. 
The one major difference from an incoming light in the TM polarization, is that $k_z^{prop}$ in the TE polarization behaves differently than in the TM polarization. Approximating the grating slits to infinite metallic slab waveguides (with a correction to the width $a$ in case of non-ideal metal to account for skin depth), $k_z^{prop}$ is then given by the equation
\begin{equation}
\label{kProp}
k_z^{prop} = \sqrt{\frac{\mu\epsilon}{c^2}\omega^2 - \gamma^2}
\end{equation}
with $\gamma = \frac{\pi m}{a}$

\begin{table}
\caption{Summary of the different cases in fig. \ref{fig:model}}
\label{table1}
\begin{tabular}{|c|c|c|c|}
\hline
Configuration & layer $n_2$ & Polarization & Spectral range \\
\hline
(1) & not present & TM & all \\
 & not present & TE & $\lambda / n_s < 2a$ \\
(2) & present & TM & all \\
 & present & TE & $\lambda / n_s < 2a$ \\
(3) & present & TE & $\lambda / n_s > 2a$ \\
\hline
\end{tabular}
\end{table}

From Eq.~\ref{kProp} it is clear that there is a cutoff frequency. Hence, in the subwavelength regime ($(\lambda / n_s) > 2a$) there are \textbf{no propagating modes} inside the grating. This seems to suggest that the mechanism for EOT is unavailable. However, EOT in wire grating experiments and simulations in the TE polarization \cite{crouse_polarization_2007,lochbihler_enhanced_2009} and in gratings with an added thin layer of dielectric material \cite{moreno_extraordinary_2006} were observed. Both these phenomena can be explained by our model. In the wire gratings the configuration was such that $2a > d$. For $(\lambda / n_s) > d$, the first bragg diffraction is evanescent before and after the grating, as in the TM polarization. As long as $(\lambda / n_s) < 2a$ (below the cutoff frequency) there will be a propagating bloch wave inside the grating, and thus, a standing wave condition can be achieved. Therefore, we expect EOT in the TE polarization as well, as long as $d < (\lambda / n_s) < 2a$, since this condition is not actually subwavelength. Thus the configuration discussed in these articles corresponds the model in Fig.~\ref{fig:model}(1b). 

For EOT to occur in the TE polarization in the subwavelength regime, the system has to be configured differently, so a standing wave will be possible. When a thin dielectric layer is added above of the grating, there is the possibility of a waveguide mode even for $(\lambda / n_s) > 2a$ \cite{moreno_extraordinary_2006} (given that $\lambda/n_2<d$). For $(\lambda / n_s) < 2a$, both the grating and the dielectric layer $n_2$ support a propagating mode in the first bragg order ($m=1$), and therefore the standing wave is given by the equation for a two layer waveguide (with $n_2$ and $n_{eff}$ for the grating layer), corresponding to Fig.~\ref{fig:model}(b)(2).

For $(\lambda / n_s) > 2a$, there is no longer a propagating mode inside the grating, while the dielectric layer still supports one. However, there will still be an eigenmode in the grating with a relatively small imaginary part, which will be denoted by $k_z^{ev}$. For thin gratings, an evanescent coupling by the first bragg diffraction to the waveguide mode in the thin dielectric layer will still be possible. As stated by Garcia Vidal et al. in Ref.~\cite{moreno_extraordinary_2006} transmission minima will appear for wavelengths approximately satisfying the waveguide condition in the dielectric layer, with a metallic slab instead of the grating, with an explanation similar to the one discussed by Rosenblatt et al. in Ref.~\cite{rosenblatt_resonant_1997}. In contrast, if we again examine the first bragg order in the grating:
\begin{equation}\label{FirstBraggTE}
E(r) = Ae^{i[(k_x+g)x + k_z^{ev}z]\hat{y}}
\end{equation}
and define $n_{eff}$ for the effective dielectric layer as shown in Eq.~\ref{neff}, it will be shown in the next section that the maximum transmission peaks  observed in this configuration satisfy the waveguide condition, with the effective dielectric layer. Furthermore, these maxima depend on $n_{eff}$ as predicted by the model, while the transmission minima are not. This corresponds to the configuration in Fig.~\ref{fig:model}(b)(3).

\section{Comparison to numerical calculations}
To check the validity of our model predictions for the spectral position of the EOT maxima, we compare them to a full numerical calculation using an RCWA method\cite{treacy_dynamical_2002, moharam_rigorous_1986}.

\begin{figure}
\includegraphics[width=80mm]{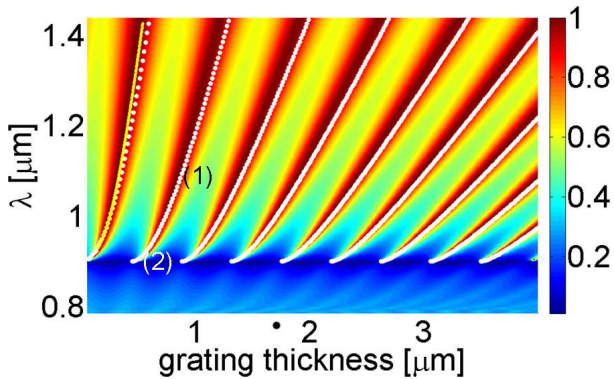}
\caption{\label{fig:TM_transmission} Transmission in the TM polarization with no added thin dielectric layer, in the symmetric configuration $n_1 = n_3 = n_s$, for different wavelengths and grating thickness. The dotted white lines are the transmission maxima according to the EBC model. The yellow line is the first transmission maximum according to the EBC model with all the diffraction orders taken into account. The periodicity is $d=0.9\mu m$ and the slit width is $a=0.35\mu m$.}
\end{figure}

Fig \ref{fig:TM_transmission} shows the numerically calculated zero order transmission intensity for different wavelengths and grating thickness . The numerical calculation was done with $d = 0.9\mu m$, $a = 0.35\mu m$, and ($n1 = n3 = n_s = 1$) on both sides of the grating and inside the slits. This corresponds to the configuration depicted in fig.~\ref{fig:model}(b)(1). The predictions of the EOT maxima, given by Eq.~\ref{EOTConditionTM} are plotted in fig.~\ref{fig:TM_transmission}, by the white dotted lines. A very good agreement between the numerically calculated transmission maxima and the EBC model is clearly seen, with no fitting parameters.

It is apparent from fig. \ref{fig:TM_transmission} that the transmission maximum occurs at different wavelengths for different grating thickness \cite{garcia-vidal_transmission_2002} as expected from a Fabri-Perot like behavior. Furthermore, there is a smooth transition between the cavity-like maximum, as indicated by region (1) in fig \ref{fig:TM_transmission} (a), in which we see the familiar linear dependence of the resonant wavelength on the cavity width $w = \frac{\lambda_0 l}{2n_s}$ ($l$ being an integer) and the SPP-like maximum, as indicated by region(2), which deviates from this slope. As seen, this transition is predicted by the EBC model. Since $n_1=n_3=n_s$, in the limit $\frac{\lambda}{n_1} \gg d$, which means $n_{eff} \gg n_{1,3}$ and $\chi\gg 1$, Eq.~\ref{phi123} becomes $\phi_{12} = \phi_{23} \approx tan^{-1}(\chi^3) \approx \pi / 2$ and Eq. ~\ref{EOTConditionTM} reduces to $2n_sk_0w = 2\pi l$ which is the metallic slab waveguide condition, or
$$w = \frac{\lambda_0 l}{2n_s}, $$ with $l \in \mathbb{N}$.
This means that the standing wave is confined exactly inside the metallic grating, corresponding to the cavity-like EOT maxima in Ref. \onlinecite{porto_transmission_1999}. The other limit, where $\frac{\lambda}{n_1} \rightarrow d $ gives us a confined mode with an effective length in the $\hat{z}$ direction which is much larger than the grating width $w$. This limit corresponds to the SPP-like maxima in Ref. \onlinecite{porto_transmission_1999}: due to the slow spatial decay of the field intensity of the confined mode into the surrounding dielectric layers in this limit, the near field intensity distribution behaves as a surface plasmon-like mode. In this sense, the cavity-like modes and the SPP-like modes are just two limits of the general EBC model.

\begin{figure}
\includegraphics[width=80mm]{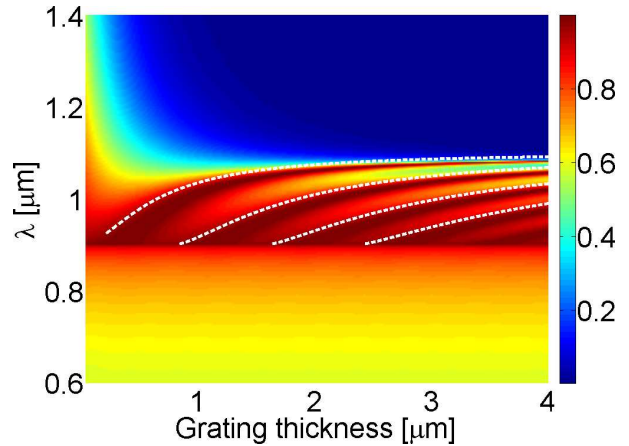}
\caption{\label{fig:TE_transmission} Transmission in the TE polarization for different wavelengths and grating thicknesses.  The dashed lines are the transmission maxima according to the EBC model . The periodicity is $d=0.9\mu m$, the slit width is $a=0.55\mu m$, so that $2a > d$ in the configuration corresponding to fig. ~\ref{fig:model}(3).}
\end{figure}

In Fig \ref{fig:TE_transmission} we see a similar graph for the TE polarization, with $d = 0.9\mu m$ and $a = 0.55\mu m$. we can see that for $(\lambda / n_s) < d$ the transmission is not dependent on the grating thickness, and no EOT resonance is observed. However, since $d < 2a$ we see EOT in this polarization when $d < (\lambda / n_s) < 2a$, again in good agreement with our model. As can be seen, the behavior of the EOT lines differs from the TM polarization. This difference is largely explained by the difference in the metallic slab waveguide equations (and thus $k_z^{prop}$) between TE and TM.

In both polarizations, the standing wave model predicts quite accurately the behavior of the EOT, but with a slight deviation from the actual maximum. This can mostly be explained by the fact that our approximation took into account only the first bragg order. The yellow line in fig.~\ref{fig:TM_transmission} is the EBC model calculated first transmission maximum when all the diffraction orders of the waveguide condition\cite{shen_mechanism_2005} are taken into account. As can be seen, this improves the accuracy of the model's prediction. Furthermore, as was previously shown \cite{garcia-vidal_transmission_2002, lalanne_one-mode_2000} taking only the propagating mode inside the grating into account already causes a small shift of the transmission features.

\begin{figure}
\includegraphics[width=80mm]{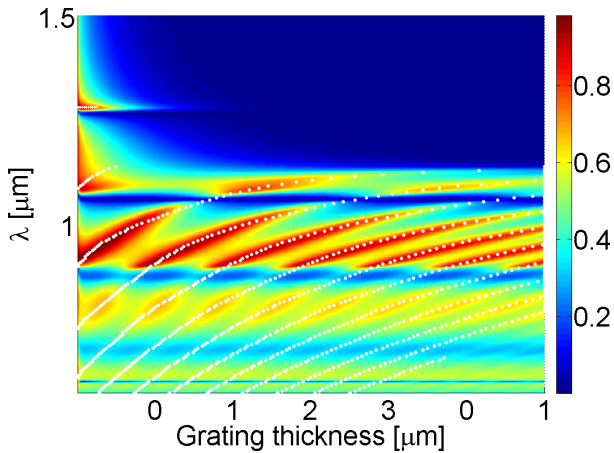}
\caption{\label{fig:TE_pol_transmission} Transmission in the TE polarization with a thin dielectric layer for different wavelengths and grating thickness. The dashed lines are the transmission maxima according to the EBC model (dotted line). The periodicity is $d=0.9\mu m$, the slit width is $a=0.35\mu m$. Because $n_2 = 1.52$ and the metal is not ideal, the cutoff is at $\lambda=1.12 \mu m$}
\end{figure}

A more interesting case is a configuration in which a thin dielectric layer is added on top of the grating (configuration (c) in fig~\ref{fig:model}. in Fig.~\ref{fig:TE_pol_transmission} we again see the transmission for different wavelengths and grating thickness, with $d = 0.9\mu m$, $a = 0.35\mu m$, $w_2=0.93\mu m$.

It is clear from fig.~\ref{fig:TE_pol_transmission} that there is a cutoff in the transmission around $\lambda=1.12\mu m$, which corresponds to the transition into the subwavelength regime. As can be seen, in the non-subwavelength regime ($(\lambda / n_s) < 2a$), the maximum transmission lines behave similarly to the case in which there is no added dielectric layer, with the difference that there is still a propagating mode in the dielectric material.

\begin{figure}[h]
\includegraphics[width=60mm]{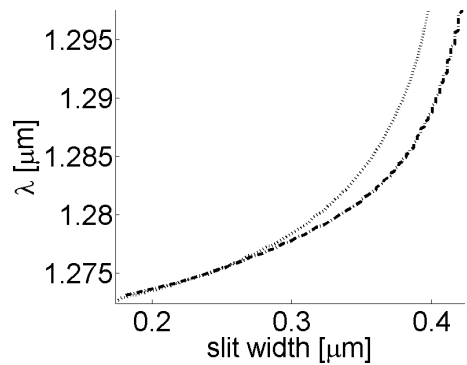}
\caption{\label{fig:TE_sub_slitwidth} Transmission maxima according to the RCWA (dashed line) and the standing wave model (dotted line) for different slit widths. The transmission minimum is at $\lambda = 1.726 \mu m$. The periodicity is $d=0.9\mu m$, the grating width is $w=0.25\mu m$ and the dielectric layer width is $w_d=0.25\mu m$.}
\end{figure}

The extra observed features are lines of minima in the transmission, which closely correspond to the waveguide condition in the thin dielectric layer $n_2$, taking the grating as a homogeneous metallic slab\cite{moreno_extraordinary_2006}, and accordingly do not change with the metal width. However, in the subwavelength regime, there is a transmission maximum near to the waveguide minimum, which does not change spectrally with the metal thickness. Because there is no propagating mode in the grating, this is to be expected, as already explained. However, we do expect it to change accordingly to the imaginary part of $k_z$ (defined as $k_z^{ev}$). Changing the imaginary part of $k_z^{ev}$ can be done by changing the slit width $a$. In Fig. ~\ref{fig:TE_sub_slitwidth} the zero order transmission maxima is extracted from the numerical model for different values of the slit width $a$ in the subwavelength regime, and is compared to the predicted value given by the EBC model. As can be seen, there is a pretty good correspondence between the two. Note that the resolution is of the magnitude of $10-20 nm$, and so we do not expect a perfect fit. However it is clear that the trend of both lines is the same.

We have also compared the EBC model to numerical calculations in which $n_1\neq n_3$. In the spectral regimes where all the higher diffraction orders are evanescent in both infinite dielectric layers (layers 1,3), the agreement with our analytical model was just as good. When one of the dielectric layers starts supporting a propagating mode with $m\neq0$, no real EOT is apparent. The model is also in good agreement with the numerical calculations when the metal is not taken as ideal but given by the drude model, with the exception that when calculating $k_z^{prop}$ in the TE polarization, the skin depth needs to be taken into account.

\section{Conclusions}
Initially, it was believed that SPPs are involved in the emergence of EOT in 1D subwavelength metallic gratings. Since then a wide range of configurations that do not support SPPs  but exhibit EOT have been reported to exist. However, it is evident that all of these configurations do support a standing wave for the higher diffraction orders when exhibiting EOT, and do not when no EOT is possible. Therefore, our conclusion is that a standing wave, which can be similar to a Fabri-Perot mode, or a waveguide mode, is always the mechanism responsible for EOT. To prove this point, we have developed a simple analytical model for the case where $\frac{a}{d} \sim 0.5$. We showed that this model is in a very good agreement with the full numerical calculations for a wide range of configurations, and therefore is a good approximation for describing the underlying physical picture for EOT. Such a simple analytical model can also be helpful in designing EOT structures for various possible applications.

\bibliography{my_bib}{}

\end{document}